\documentclass{ws-ijmpd}

\begin{document}
\markboth{Hao and Akhoury} {A Possible Late Time $\Lambda$CDM-like
Background Cosmology in Relativistic MOND Theory}

\title{A Possible Late Time $\Lambda$CDM-like
Background Cosmology in Relativistic MOND Theory}
\author{J. G. Hao}
\address{Department of Physics, The University of Michigan, Ann Arbor, MI 48109,
USA \\jghao@umich.edu}
\author{R. Akhoury}
\address{Michigan Center For Theoretical Physics, Department of Physics,
The University of Michigan, Ann Arbor, MI 48109, USA
\\ akhoury@umich.edu}

\maketitle

\date{\today}

\begin{history}
\received{Day Month Year}
\revised{Day Month Year}
\comby{Managing
Editor}
\end{history}

\begin{abstract}
In the framework of Relativistic MOND theory (TeVeS), we show that a
late time background $\Lambda$CDM cosmology can be attained by
choosing a specific $F(\mu)$ that also meets the requirement for the
existence of Newtonian and MOND limits. We investigate the dynamics
of the scalar field $\phi$ under our chosen $F(\mu)$ and show that
the "slow roll" regime of $\phi$ corresponds to a dynamical
attractor, where the whole system reduces to $\Lambda$CDM cosmology.
\end{abstract}


\section{Introduction}
Current astronomical observations on cosmological scales (CMB
anisotropy \cite{spergel,bennett}, supernovae
\cite{riess,perl,tonry} and SDSS\cite{SDSS}) reveal that our
universe is spatially flat, with about two thirds of the energy
content resulting from what is referred to as dark energy. This
energy
 has a  negative pressure to account for the accelerated expansion of the
 universe. While, on smaller, galactic scales, the rotational curves
of galaxies strongly indicate that the biggest contributions to
their mass density  arises from non-luminous matter, which has
given rise to speculations on the existence of dark
matter\cite{dark matter}.

On the other hand, it is also interesting to inquire if we can solve
the current observational puzzles, both dark matter and dark energy,
by modifying Einstein gravity in spite of  its many successes, in
particular, the solar system tests. MOND theory is a striking
modification of the Newtonian laws of motion and explains the
rotational curves with amazing accuracy\cite{MOND} without
introducing any dark matter. After some initial problems, a
consistent relativistic extension of the MOND theory has recently
been proposed by Bekenstein\cite{bekenstein}, which has spurred
extensive studies on its successes and
pitfalls\cite{skordis,silk,sanders,zhao1,zhao2,chiu,famaey,ratranew,bourliot}.
In that extension, general relativity has been modified by including
two scalar fields (one dynamical while the other is non-dynamical),
one vector field. In addition, the theory contains one arbitrary
function whose form is dictated in part by its primary purpose,
i.e., to address the data on the rotational curves of galaxies, or
equivalently, the dark matter problem. Consistency with data from
the Bullet clusters\cite{clowe} indicates that eV scale hot
neutrinos must also be included as explicit dark matter
candidate\cite{angus}. The remarkable features of this theory are
that it can explain the galaxy rotational curves without introducing
dark matter (just as its non-relativistic counterpart, MOND theory)
and at the same time reduce to general relativity in the appropriate
limits. The price one pays is the additional
 complication of the whole system: two scalar fields, one vector
 field and gravitational field.

A natural question to ask is if the new framework also works for
cosmology? Can it help to tackle the dark energy problem? We will
show in this paper that under appropriate conditions, the scalar
field could mimic the behavior of the dark energy
models\cite{ratra,Steinhardt,Peebles,mukhanov,caldwell}.
Specifically, in the "slow roll" regime, it will be seen to reduce
to the standard $\Lambda$CDM model. Although the current formulation
of TeVeS is still not as mature as the dark matter models, it has
the potential, to naturally solve the cosmic puzzles currently
explained by dark energy and dark matter without invoking either of
them explicitly\cite{bekenstein_rev}.

\section{Basic setup}

We begin with a brief outline of Gravity in the TeVeS framework.
Following Bekenstein\cite{bekenstein}, the action for the whole
system could be written as
\begin{equation}\label{action}
S=S_g+S_s+S_v+S_m
\end{equation}

\noindent where
\begin{eqnarray}\label{actiong}
S_g=(16\pi G)^{-1}\int g^{\alpha\beta} R_{\alpha\beta} \sqrt{-g}
d^4 x
\end{eqnarray}

\begin{equation}\label{actions}
S_s =-{\frac{1}{2}}\int\big[\sigma^2
h^{\alpha\beta}\phi_{,\alpha}\phi_{,\beta}+{\scriptstyle
1\over\scriptstyle 2}G \ell^{-2}\sigma^4 F(kG\sigma^2)
\big]\sqrt{-g} d^4 x
\end{equation}

\begin{eqnarray}\label{actionv}
S_v =-{K\over 32\pi G}\int \big[g^{\alpha\beta}g^{\mu\nu}
\mathfrak{U}_{[\alpha,\mu]} \mathfrak{U}_{[\beta,\nu]}
-2(\lambda/K)(g^{\mu\nu}\mathfrak{U}_\mu \mathfrak{U}_\nu
+1)\big]\sqrt{-g} d^4 x,
\end{eqnarray}


\begin{equation}\label{actionm}
S_m=\int \mathcal{L}(\tilde g_{\mu\nu}, f^\alpha, f^\alpha_{|\mu},
\, \cdots)\sqrt{-\tilde g} d^4x
\end{equation}

\noindent are the actions for gravity, scalar field, vector field
and luminous matter respectively. In the framework of TeVeS, there
are two different frames, physical frame corresponding to the
"real universe" and Einstein frame. The metric in physical frame
($\tilde g_{\mu\nu}$) relates that in Einstein frame
($g_{\mu\nu}$) by $\tilde g_{\alpha\beta} = e^{-2\phi}
(g_{\alpha\beta}+\mathfrak{U}_\alpha \mathfrak{U}_\beta) -
e^{2\phi}\mathfrak{U}_\alpha \mathfrak{U}_\beta$. Here
$\mathfrak{U}_\alpha$ is a timelike 4-vector field, $\phi$ and
$\sigma$ are respectively the dynamical and the non-dynamical
scalar fields, F is a free function to be specified by dynamics
while $k$ and $K$ are two positive dimensionless parameters.
Varying the action, one can arrive at the following equations of
motion for the vector, scalar and gravitational fields
\cite{bekenstein}:

\begin{eqnarray} &K&
\left(\mathfrak{U}^{[\alpha;\beta]}{}_{;\beta}+\mathfrak{U}^\alpha
\mathfrak{U}_\gamma
\mathfrak{U}^{[\gamma;\beta]}{}_{;\beta}\right)+ 8\pi G\sigma^2
\big[\mathfrak{U}^\beta\phi_{,\beta}\,g^{\alpha\gamma}\phi_{,\gamma}+
\mathfrak{U}^\alpha (\mathfrak{U}^\beta\phi_{,\beta})^2\big]
\nonumber
 \\ &=&8\pi G (1-e^{-4\phi})
\big[g^{\alpha\mu} \mathfrak{U}^\beta  \tilde T_{\mu\beta} +
\mathfrak{U}^\alpha \mathfrak{U}^\beta \mathfrak{U}^\gamma \tilde
T_{\gamma\beta}\big] \label{vectoreq2}
\end{eqnarray}

\begin{equation}\label{sigma eq}
-\mu
F(\mu)-\frac{1}{2}\mu^2F'(\mu)=k\ell^2h^{\alpha\beta}\phi_{,\alpha}\phi_{,\beta}
\end{equation}

\begin{equation}
\left[\mu\left(k\ell^2 h^{\mu\nu}\phi_{,\mu}\phi_{,\nu}\right)
h^{\alpha\beta}\phi_{,\alpha} \right]_{;\beta}=
kG\big[g^{\alpha\beta} + (1+e^{-4\phi}) \mathfrak{U}^\alpha
\mathfrak{U}^\beta\big] \tilde T_{\alpha\beta}. \label{s_equation}
\end{equation}

\begin{equation} G_{\alpha\beta} = 8\pi G\Big[\tilde
T_{\alpha\beta} +(1-e^{-4\phi}) \mathfrak{U}^\mu \tilde
T_{\mu(\alpha} \mathfrak{U}_{\beta)} +\tau_{\alpha\beta}\Big]+
\Theta_{\alpha\beta} \label{gravitationeq}
\end{equation}

\noindent where
\begin{eqnarray}
\tau_{\alpha\beta}&\equiv&
\sigma^2\Big[\phi_{,\alpha}\phi_{,\beta}-{\frac{1}{2}}g^{\mu\nu}\phi_{,\mu}\phi_{,\nu}\,g_{\alpha\beta}-
\mathfrak{U}^\mu\phi_{,\mu}\big(\mathfrak{U}_{(\alpha}\phi_{,\beta)}-
{\frac{1}{2}}\mathfrak{U}^\nu\phi_{,\nu}\,g_{\alpha\beta}\big)\Big]
\label{tau} \nonumber
\\ &-&{\frac{1}{4}}G
\ell^{-2}\sigma^4 F(kG\sigma^2)  g_{\alpha\beta}
\end{eqnarray}
\begin{equation}
\Theta_{\alpha\beta}\equiv
K\Big(g^{\mu\nu}\mathfrak{U}_{[\mu,\alpha]}
\mathfrak{U}_{[\nu,\beta]} -\frac{1}{4}
g^{\sigma\tau}g^{\mu\nu}\mathfrak{U}_{[\sigma,\mu]}
\mathfrak{U}_{[\tau,\nu]}\,g_{\alpha\beta}\Big)- \lambda
\mathfrak{U}_\alpha\mathfrak{U}_\beta \label{Theta}
\end{equation}

\noindent $\tilde T_{\mu\nu}$ is the energy momentum tensor of
ordinary matter in the physical coordinate system.
$h^{\alpha\beta}=g^{\alpha\beta}-\mathfrak{U}^{\alpha}\mathfrak{U}^{\beta}$
and $\mu=kG\sigma^2$.

\section{Cosmological dynamics in Einstein frame}

In a cosmological setting, the symmetries of the FRW universe will
simplify the above equations considerably. It is worth recalling
here an underlying assumption in such applications. One interprets
the fields in the problem, say the scalar field, as consisting of
both a spatially homogenous (time dependent) part and an
inhomogeneous one. On cosmological scales the spatially
inhomogeneous part may be neglected in the first approximation
while at galactic scales, the inhomogeneous part plays a prominent
role and the homogenous part is negligible (quasi-static
approximation). Henceforth, we will restrict ourselves to
cosmological  scales, more specifically to the flat universe case,
and consider the line element:

\begin{equation}\label{line element}
d\tilde s^2=-d\tilde t^2+\tilde a(\tilde t)^2 d\textbf{x}^2
\end{equation}

\noindent where $d\tilde t=e^{\phi}dt$ and $\tilde a=e^{-\phi}a$. It
is clear that the physical frame and Einstein frame become the same
once the $e^{\phi}$ is constant 1. The vector field in this case
could be chosen as $\mathfrak{U}^{\mu}= \delta_t^{\mu}$ and the
scalar field depends only on time $t$. In Einstein frame, the
equations of motion and the field equations have simpler forms
though their dynamical evolution are essentially the same as those
in physical frame (they are related by a non-singular
transformation). So, in this section, we will analyze the dynamics
of the system in Einstein frame.

\begin{equation}\label{friedman eq}
\bigg(\frac{\dot{a}}{a}\bigg)^2=\frac{8 \pi
G}{3}(\rho_M+\rho_{\phi})
\end{equation}

\begin{equation}\label{acceleration eq}
\frac{\ddot{a}}{a}=-\frac{4 \pi
G}{3}(\rho_M+\rho_{\phi}+3p_M+3p_{\phi})
\end{equation}

\noindent where $\rho_M=\tilde{\rho} e^{-2\phi}$, $p_M=\tilde {p}
e^{-2\phi}$ are respectively the density and pressure of
non-relativistic matter. $\rho_{\phi}$ and $p_{\phi}$ are the
effective energy density and pressure of the scalar field $\phi$,
defined as:

\begin{eqnarray}\label{rho and p}
\rho_{\phi}=\frac{\mu \dot{\phi}^2}{k G}+\frac{\mu^2}{4\ell^2k^2
G}F(\mu)=\frac{\mu^2[3F(\mu)+\mu F'(\mu)]}{4\ell^2k^2 G}\\\nonumber
p_{\phi}=\frac{\mu \dot{\phi}^2}{k G}-\frac{\mu^2}{4\ell^2k^2
G}F(\mu)=\frac{\mu^2[F(\mu)+\mu F'(\mu)]}{4\ell^2k^2 G}
\end{eqnarray}

\noindent where $\mu$ relates to $\dot{\phi}$ via the $\sigma$
equation (Eq.\ref{sigma eq}):

\begin{equation}\label{sigma relation}
\mu F(\mu)+\frac{1}{2}\mu^2 F'(\mu)=2k\ell^2\dot{\phi}^2
\end{equation}

\noindent Here we need to emphasize, as suggested in
\cite{ratranew}, that the above definitions are held in the
effective sense due to the fact that $\rho_M$ and $p_M$ terms
contain $\phi$. But in the "slow roll" regime ($\dot{\phi} \sim 0$
as we will see later), they are well defined as the $\phi$ is merely
a constant factor. The equation of motion of the scalar field $\phi$
in a FRW background is given by

\begin{equation}\label{scalar}
\ddot{\phi}+(3H+\frac{\dot{\mu}}{\mu})\dot{\phi}+\frac{ k
G}{2\mu}(\rho_M+3p_M)=0
\end{equation}

\noindent where $H=\dot{a}/a$ and dot denotes the derivative with
respect to $t$. It is worth nothing that, under the specific form of
$F(\mu)$ we choose in Eq.(\ref{bfform}),  Eq.(\ref{scalar}) is
consistent with the "slow roll" approximation $\dot{\phi}\sim 0$
under two different situations: First, as in the usual scalar field
dark energy dynamical system, the matter density $\rho_M$ decreases
to zero at late times, which leads to a $\rho_{\phi}$ dominated
universe; Second, $\frac{k}{\mu} \ll 1$, which lead to a universe
with both $\rho_M$ and $\rho_{\phi}$. In next section, we consider
the second possibility when fitting the model to the Supernovae
data. On the other hand, if we assume that $\dot{\phi}$ varies
slowly, then we can consider the two energy components to be
approximately adiabatic, and energy momentum conservation leads to

\begin{equation}\label{conservation}
\frac{d\rho_i}{dt}=-3 \frac{\dot{a}}{a}(\rho_i+p_i)
\end{equation}

\noindent where the subscript $i$ denotes $M$ and $\phi$. To get a
closed system of equations, we also need to specify the equation
of state of the energy/matter content. This equation of state is
specified by $w=p/\rho$, which is 0 for non-relativistic matter.
While for the scalar field, this equation of state is given by its
equation of motion, i.e.,Eq.(\ref{scalar}) which can be expressed as,

\begin{equation}\label{eofs scalar}
w_{\phi}=\frac{4\ell^2k\mu\dot{\phi}^2-\mu^2F(\mu)}{4\ell^2k\mu\dot{\phi}^2+\mu^2F(\mu)}=\frac{F(\mu)+\mu
F'(\mu)}{3F(\mu)+\mu F'(\mu)}
\end{equation}

\noindent In the "slow roll" regime, $\dot{\phi}\sim\ddot{\phi}\sim
0$ and $w_{\phi}\sim -1$. It is worth noting that from Eq.(\ref{eofs
scalar}), the equation of state could approach $-1$ from either
greater than -1 (quintessence case) or less than -1(phantom case) if
we choose the form of $F(\mu)$ appropriately. In this paper, we
focus on the case with $w_{\phi}\geq -1$. Eqs.(\ref{friedman
eq}-\ref{acceleration eq}) are expressed in terms of the scale
factor in the Einstein picture. In the following, we will continue
to work in Einstein frame to investigate the dynamics of the scalar
field.

From the above discussion, it is clear now that we will get the
evolution of the whole dynamical system if the form of $F(\mu)$ is
specified. However, at present there are no theoretical arguments
in favor of any specific choice, thereby providing a lot of
freedom in this regard. The constraints on the form of $F(\mu)$
are phenomenological in nature motivated by the condition that the
correct physics is obtained. Besides the toy model $F(\mu)$
suggested by Bekenstein, there are some others forms suggested to
account for the better fit to data\cite{zhao2,famaey}. In this
section, we will consider the the choice of the form of $F(\mu)$
that will lead to an acceptable cosmology in addition to
accommodating the MOND theory and Newton's Law. From Eq.(\ref{rho
and p}), the positive energy condition, $\rho_{\phi}\geq 0$ and
the non-vanishing of $\rho_{\phi}$ as $\dot{\phi}\rightarrow 0$
may restrict the freedom in the choice of the form of $F(\mu)$. On
the other hand, the existence of MOND and Newtonian limits also
give constraints on the form of $F(\mu)$ as\cite{bourliot}

\begin{equation}\label{Fconstrant}
\frac{d(\mu^2F(\mu))}{d
\mu}=-\frac{k^2\mu^2}{64\pi^2(1-\mu)^m}f(\mu)
\end{equation}

\noindent with $f(\mu)$ an arbitrary function of $\mu$ with the
requirement of non-vanishing $f(0)$ and $m=1$ is the case proposed
by Bekenstein\cite{bekenstein}. Therefore, we suggest a form of
$F(\mu)$ that meets the above requirements as
\begin{equation}\label{bfform}
F(\mu)=\frac{3}{8}\frac{\mu(4+2\mu-4\mu^2+\mu^3)+2\ln[(1-\mu)^2]}{\mu^2}+\frac{\alpha}{\mu^2}
\end{equation}

\noindent where $\alpha$ is a dimensionless constant. When
$\alpha=0$, the above reduces to the form in\cite{bekenstein}.
Then, from Eq.(\ref{sigma relation}), we can obtain the relation
between $\mu$ and $\dot{\phi}$ as
\begin{equation}\label{muphirelation}
\frac{3}{2}(1+\mu-3\mu^2+\mu^3-\frac{1}{1-\mu})=4
k\ell^2\dot{\phi}^2
\end{equation}

\noindent For a consistent cosmology, we require
$2\leq\mu<\infty$. Note also that $\dot{\phi}=0$ when $\mu=2$, for
which the energy density Eq.(\ref{rho and p}) becomes
$\rho_{\phi}=\frac{\alpha}{4\ell^2k^2 G}$ and the equation of
state reduces to,  $w_{\phi}=-1$ at late times, which corresponds
to a de Sitter phase. This not only provides an acceptable
cosmological state but also ensures the consistency of our
previous assumption $\dot{\phi}\sim 0$. Next, we show that such a
phase corresponds to a dynamical de Sitter attractor\cite{hao}. To
do this, we introduce the dimensionless variables $x=\phi$,
$z=t_0\dot{\phi}$ and $N=\ln a$ with $t_0$ a scaling constant of
dimension $t$. We can rewrite the Eqs.(\ref{friedman
eq}-\ref{conservation}) in terms of $x$, $z$ and $N$ and linearize
them around the critical point $(x, z)=(x,0)$, then, we arrive at
the following system of equations

\begin{eqnarray}\label{neweq}
\frac{dx}{dN}&=&\frac{z}{t_0\sqrt{2\pi\alpha/3k^2\ell^2}}\\\nonumber
\frac{dz}{dN}&=&-3z
\end{eqnarray}

\noindent It is easy to see that the eigenvalues of the
coefficients matrix of Eq.(\ref{neweq}) are $(-3,0)$ which
indicates that the critical point is stable, i.e. a dynamical
attractor. One comes to the same conclusion by noting that
$\dot{\phi}=0$ leads to a minimum of the energy density
$\rho_{\phi}$. In Figs. ( 1 ) and ( 2 ), we plot the numerical
results for the dynamical system defined above. Note that our
intention here is merely to illustrate the possibility of a
consistent cosmology and not to fit the exact observational data.
So, for convenience, we have set the parameter $\alpha=0.01$ and
the rest to unity. Since the attractor corresponds to all $\phi$
with $\dot{\phi}=0$, we can choose $\phi \sim 0$ so that
$e^{-2\phi}\sim 1$, which is indicated by the previous discussion.
In our numerical analysis, we chose $\phi=0.00001$ and the initial
$\dot{\phi}$ from 0.001 to 0.005 with an increment of 0.001. We
should point out that the above conclusion won't change even if we
choose other values for $\phi$. Then the physical frame and
Einstein frame are different by a factor of $e^{-2\phi}$. In Fig.
(3), we plot the corresponding evolution of the $\rho_{\phi},
p_{\phi}$ and $w_{\phi}$ with respect to $\mu$ given the form of
$F(\mu)$ in Eq. (\ref{bfform}).

\begin{figure}
\begin{center}
\epsfig{file=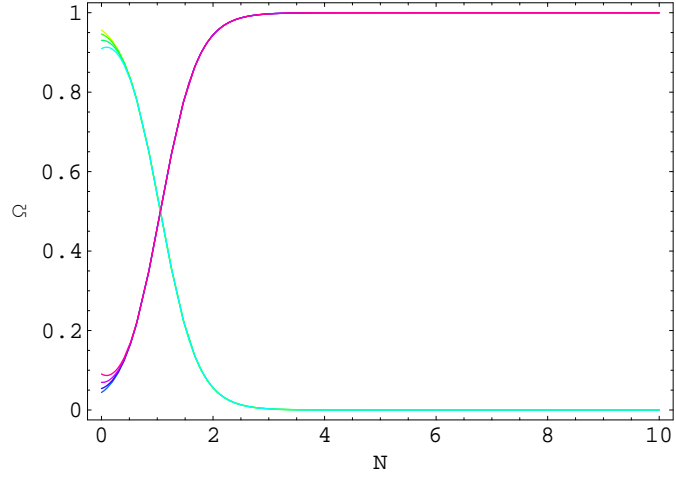,height=2.5in,width=3.5in} \caption{The
evolution of cosmic parameters for matter(indigo curve) and $\phi$
field (pink curve). From the plots, one can find the curves with
different initial conditions converges very quickly to a common
track corresponding to the attractor.}
\end{center}
\end{figure}

\begin{figure}
\begin{center}
\epsfig{file=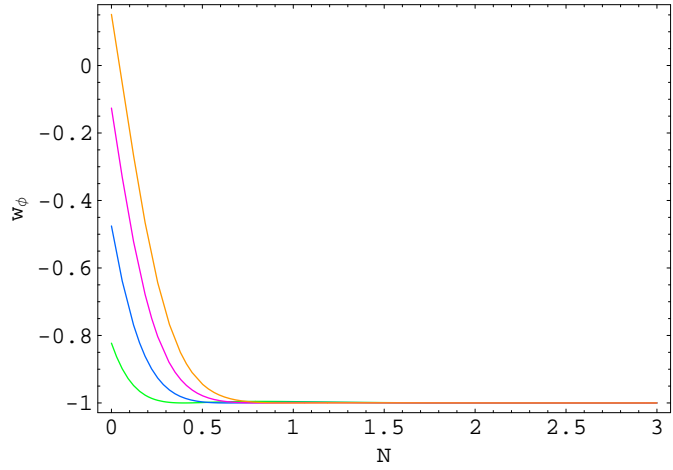,height=2.5in,width=3.5in} \caption{The
evolution of the equation of state of $\phi$ field for different
initial $\dot{\phi}$.}
\end{center}
\end{figure}

\begin{figure}
\begin{center}
\epsfig{file=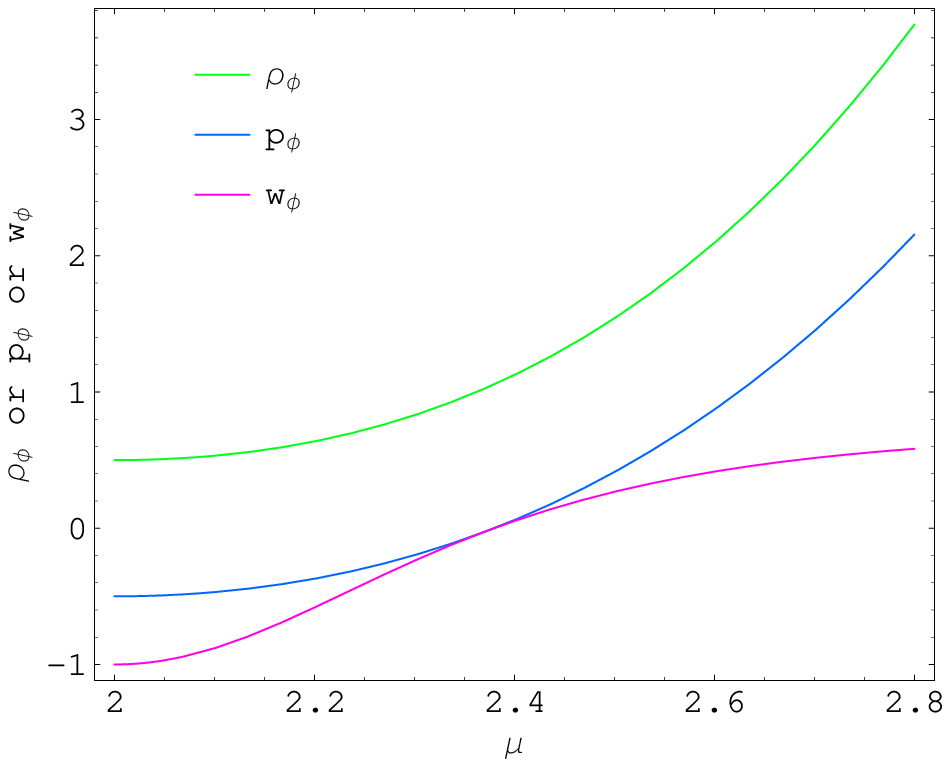,height=2.5in,width=3.5in} \caption{The
evolution of $\rho_{\phi}, p_{\phi}$ and  $w_{\phi}$. Note that
the unit for $\rho_{\phi}$ and $p_{\phi}$ are $1/4\ell^2k^2 G$
although they are plotted together with $w_{\phi}$.}
\end{center}
\end{figure}

\section{Cosmology in physical frame}
The analysis in preceding section is carried out in the Einstein
frame. But we need to work in physical frame if we want to compare
the theory with observation data. The two frames are mathematically
equivalent to each other although the physical interpretations are
different. With a simple transformation, one could convert the the
equations from Einstein frame to physical frame. Under the "slow
roll" approximation($\ddot{\phi} \sim \dot{\phi} \sim 0$), the
system can be casted as:

\begin{equation}\label{neweq1}
(\frac{\tilde a '}{\tilde a})^2=\frac{8 \pi G}{3}(\tilde \rho
e^{-4\phi}+\rho_{\phi}e^{-2\phi})
\end{equation}

\begin{equation}\label{neweq2}
\frac{\tilde a ''}{\tilde a}=-\frac{4 \pi G}{3}(\tilde \rho
e^{-4\phi}-2\rho_{\phi}e^{-2\phi})
\end{equation}

\begin{equation}\label{neweq3}
\frac{d\rho_i}{d \tilde t}=-3\frac{\tilde a '}{\tilde a}(\rho_i+p_i)
\end{equation}

\noindent Where the prime denotes derivative with respect to $\tilde
t$ and we have used $p_{\phi}=-\rho_{\phi}$ and $e^{-2\phi} \sim
const$ under the condition $\dot{\phi} \sim 0$.

In the spirit of MOND, $\tilde \rho$ is the density of baryons.
Comparing Eq.(\ref{neweq1}) to the standard $\Lambda$CDM cosmology,
we can find that the term $\tilde \rho e^{-4\phi}$ plays the
equivalent role of non-relativistic matter (baryons and dark
matter). That is to say, $\tilde \rho e^{-4\phi}-\tilde \rho=\tilde
\rho (1-e^{-4\phi})$ can be considered as an "effective" dark matter
density. We introduce $\Omega_b=\tilde \rho/\rho_{crit}$,
$\Omega_{\phi}=\rho_{\phi}e^{-2\phi}/\rho_{crit}$ and
$\Omega_M=\tilde \rho e^{-4\phi}/\rho_{crit}$ as the energy density
fraction of baryon, dark energy and non-relativistic matter
respectively, where $\rho_{crit}=3H^2/8\pi G$ is the critical energy
density. From Eq.(\ref{sigma relation}), one can observe that for
any choice of $F(\mu)$, $\mu$ is a constant as $\dot{\phi} \sim 0$
and therefore $\rho_{\phi}$ is constant in the "slow roll" regime.
As a result, the above system could be reduced to $\Lambda$CDM
universe with the relation
\begin{equation}\label{relation}
\phi=\frac{1}{4}\ln(\frac{\Omega_b}{\Omega_M})
\end{equation}

\noindent From the discussion in the previous section, we know that
$\dot{\phi}=0$ with $\phi$ at any value is a dynamical attractor.
From Eq.(\ref{relation}), we can fix the $\phi$ value via
observation, i.e. the ratio of $\frac{\Omega_b}{\Omega_M}$. From Big
Bang Nucleosynthesis\cite{liddle}, we can give a constraint on the
baryon density ($\tilde \rho$ and hence $\Omega_b$), while from
SNIa\cite{riess}, we can constrain the total density of
non-relativistic matter($\tilde \rho e^{-4\phi}$ and hence
$\Omega_M$). Combining the two, we can give a constraint on $\phi$
as $\phi=-0.46\pm 0.03$ and the results are shown in Fig.(4).

\begin{figure}
\epsfig{file=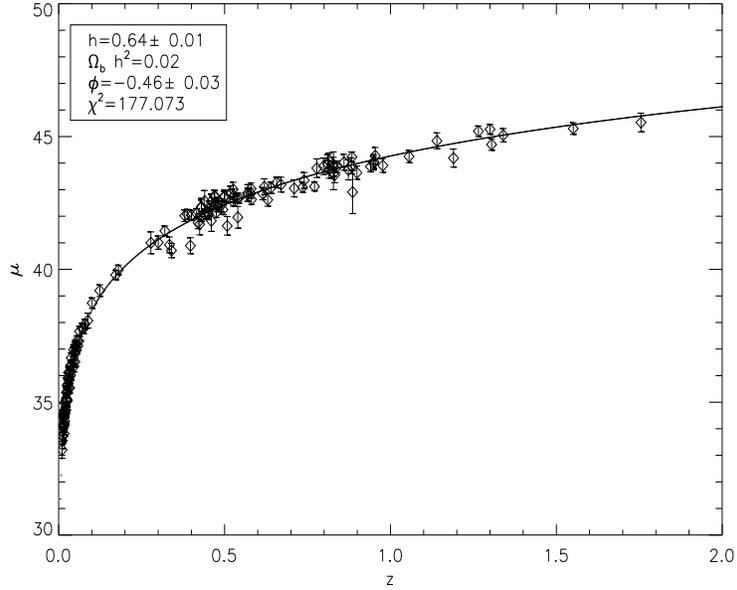,height=6.5in,width=4.5in} \caption{Constraint
on $\phi$ from Gold Sample SNIa data. The $\mu$ axis in this plot is
the distance modulus.}
\end{figure}

\section{Discussion}

In this paper, we show a possible late time background $\Lambda$CDM
like cosmology in the framework of TeVeS theory by choosing the
$F(\mu)$ as a specific form that also meets the requirements of
Newtonian and MOND limits. We show the dynamical system has a late
time de Sitter behavior and the scalar field at the de Sitter
attractor should be $-0.46 \pm 0.03$ so as to be in agreement with
current astronomical observations.

On the other hand, the framework of the TeVeS theory involves an
arbitrary function in addition to many parameters as well as
auxiliary fields. This not only provides much freedom in matching
the theory to observations, but also adds uncertainties. Clearly,
they must be constrained from further phenomenological
investigations of the testable predications of the theory. It
would be interesting to see if the model can be made compatible
with all the astronomical data with a unique choice for the
arbitrary function. Are there any theoretical constraints on this
function?

The analysis in this paper is aimed at suggesting a possibility to
reduce the complicated dynamical system of TeVeS into the customary
$\Lambda$CDM under slow roll approximation and thus get some
feedback from the cosmological observational constraints on
$\Omega_b/\Omega_M$. The choice of the function $F(\mu)$, though
without any deep theoretical basis, phenomenologically explains both
dark matter and dark energy in the same framework. It remains to be
checked in future how this form of the function $F(\mu)$ can be
embedded in the larger context of an appropriate function that
explains all the data.

\vspace{0.8cm} \noindent ACKNOWLEDGEMENT: This work was supported
in part by the US Department of Energy.


\end{document}